\documentclass[aps,prx,preprint,groupedaddress,amssymb,amsmath,floatfix,longbibliography]{revtex4-1} % w/o pacs

\usepackage{graphicx}% Include figure files
\usepackage{bm}% bold math
\usepackage{cleveref}

\newcommand{\moire}{moir\'e\:}
\newcommand{\Moire}{Moir\'e\:}

\newcommand{\WS}{WS$_{2}$\:}
\newcommand{\MoSe}{MoSe$_{2}$\:}

\usepackage{subfig}
\usepackage{color}

\begin{document}
%\title{\Moire Polaritons in van der Waals Heterobilayers}
%\title{Van der Waals Heterostructure Polaritons with \Moire-Lattice Induced Nonlinearity}
\title{Van der Waals Heterostructure Polaritons with \Moire-Induced Nonlinearity}
%\title{Moiré Polaritons and Nonlinearity in a Van der Waals Heterostructure}

\author{Long Zhang$^{1,2}$, Fengcheng Wu$^3$, Shaocong Hou$^4$, Zhe Zhang$^{1}$, Yu-Hsun Chou$^{5}$, Kenji Watanabe$^6$, Takashi Taniguchi$^7$, Stephen R. Forrest$^{1,4}$}
\author{Hui~Deng$^{1}$} \email{dengh@umich.edu}
\small{ }
\address{$^1$ Physics Department, University of Michigan, 450 Church Street, Ann Arbor, MI 48109-2122, USA}
\address{$^2$ Department of Physics, Xiamen University, Xiamen 361005, Fujian, China}
\address{$^3$ Condensed Matter Theory Center and Joint Quantum Institute, Department of Physics, University of Maryland, College Park, Maryland 20742, USA}
\address{$^4$ Department of Electrical Engineering and Computer Science, University of Michigan, 450 Church Street, Ann Arbor, MI 48109-1040, USA}
\address{$^5$ Department of Photonics, National Cheng Kung University, No. 1, University Road, East District, Tainan City 70101, Taiwan, ROC}
\address{$^6$ Research Center for Functional Materials, National Institute for Materials Science, Tsukuba, Japan}
\address{$^7$ International Center for Materials Nanoarchitectonics, National Institute for Materials Science, Tsukuba, Japan}
\maketitle

%\begin{abstract}
Controlling matter-light interactions with cavities is of fundamental importance in modern science and technology \cite{walther_Cavity_2006}. It is exemplified in the strong-coupling regime, where matter-light hybrid modes form, with properties controllable via the photon component on the optical-wavelength scale \cite{deng_exciton-polariton_2010,schneider_exciton-polariton_2016}. In contrast, matter excitations on the nanometer scale are harder to access.
In two-dimensional van der Waals heterostructures, a tunable \moire lattice potential for electronic excitations may form \cite{Bistritzer_Moire_2011}, enabling correlated electron gases in lattice potentials \cite{dean_Hofstadter_2013, cao_Unconventional_2018, tang_Simulation_2020,regan_Mott_2020a,shimazaki_Strongly_2020}. Excitons confined in \moire lattices \cite{wu_topological_2017,yu_moire_2017} have also been reported, but cooperative effects have been elusive and interactions with light have remained perturbative \cite{jin_Observation_2019,tran_Evidence_2019,seyler_Signatures_2019,alexeev_Resonantly_2019a}.
Here, integrating \MoSe-\WS heterobilayers in a microcavity, we establish cooperative coupling between \moire-lattice excitons and microcavity photons up to liquid-nitrogen temperature, thereby integrating into one platform versatile control over both matter and light. The density dependence of the \moire polaritons reveals strong nonlinearity due to exciton blockade, suppressed exciton energy shift, and suppressed excitation-induced dephasing, all of which are consistent with the quantum-confined nature of the \moire excitons.
Such a \moire polariton system combines strong nonlinearity and microscopic-scale tuning of matter excitations with the power of cavity engineering and long range coherence of light,
providing a new platform for collective phenomena from tunable arrays of quantum emitters.
%\end{abstract}


%\maketitle
\section{Introduction}
Controlling matter-light interactions has mostly been implemented with either microscopic, individual quantum particles, or macroscopic ensembles of free particles that often can be modelled classically. Bridging the two limits to realize collective coupling among arrays of quantum particles ushers new paradigms in quantum many-body physics and quantum simulation research, such as been pursued using atomic optical lattices and cavities \cite{ritsch_cold_2013,gonzalez-tudela_subwavelength_2015,leonard_Supersolid_2017}.
In solid state systems, however, such a pursuit has been exceedingly difficult, for nonlinearity is often too weak, due to screened Coulomb interactions, and inhomogeneity, too large. In this work, we show that \moire lattices formed in hetero-bilayer (hBL) transition metal dichalcogenides (TMDCs) crystals may provide a platform that overcomes these limitations of conventional solids.

When two monolayer (ML) crystals are placed together, a \moire lattice can form due to a slight mismatch in the lattice constants or crystal orientations of the MLs \cite{Bistritzer_Moire_2011,wu_topological_2017,yu_moire_2017}. Its period is tunable by the twist angle between the ML crystals from a few to a few tens of nanometers -- comparable to the range of Coulomb interactions in semiconductors. The natural formation of the \moire lattice promises the prospect of uniformity among different \moire cells across the lattice, thereby potentially enabling a new solid-state platform for cooperative phenomena between light and arrays of nonlinear quantum particles.
While exciton states induced by the \moire lattice have been reported \cite{jin_Observation_2019,tran_Evidence_2019,seyler_Signatures_2019,alexeev_Resonantly_2019a}, the fundamental question remains whether a sufficiently uniform lattice of quantized excitations can be formed in a \moire system. Exciton-light interactions have remained in the perturbative regime, and cooperative phenomena have not been reported.

\section{\Moire polaritons in \WS-\MoSe hetero-bilayers}
\begin{figure*}[t]
	\includegraphics[width=\linewidth]{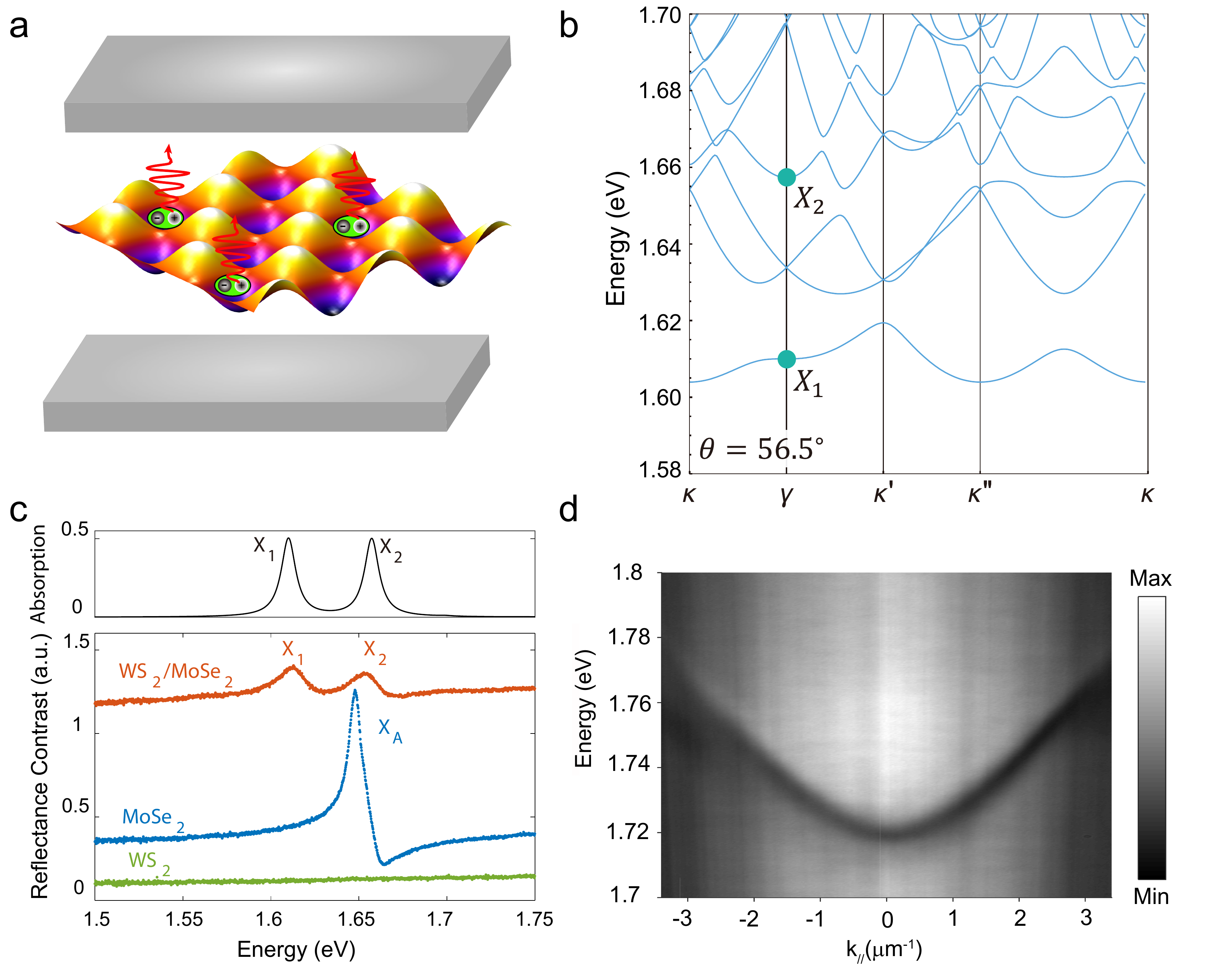}
	\caption{{\bfseries Illustration of the \moire polariton system and the constituent \moire excitons and microcavity.}
$\bold{a}$, Schematic of the \moire polariton system formed by excitons confined in \moire lattice and coupled with a planar cavity.
$\bold{b}$, moir\'e band structure for hBL excitons, where X$_1$ and X$_2$ are two bright \moire exciton states. See Method (Theory of \moire excitons) for details of the calculation.
$\bold{c}$, Top panel: theoretical optical absorption spectrum calculated from the \moire exciton band structure. Bottom panel: Reflection contrast spectra near the \MoSe A exciton resonance, from the \WS-\MoSe hBL (red), \MoSe ML (blue) and \WS ML (green). The spectra are displaced vertically for easier reading. The \MoSe ML A-exciton (X$_\text{A}$) splits into two well resolved \moire exciton (X$_1$ and X$_2$) in the hBN.
$\bold{d}$ Angle resolved white light reflection spectrum of the bare cavity, measured in a region where there is no hBL, showing a single cavity dispersion with a half linewidth $\gamma_{c}=2.7$~meV.}
    \label{fig:Schematic}
\end{figure*}

To enable cooperative coupling between \moire excitons and photons, we use \WS-\MoSe hBLs enclosed in a microcavity (Fig.~\ref{fig:Schematic}a). The hBL is  encapsulated in hexagon boron nitride (hBN) with a twist angle measured to be $56.5^{\circ} \pm 0.9^{\circ}$ (Extended Data Fig.1). The \WS-\MoSe hBL is unique among commonly studied TMDCs hBLs in that the two lowest-energy \moire exciton modes have large oscillator strengths, inherited from that of the ML \MoSe A exciton.\cite{alexeev_Resonantly_2019a,zhang_Twistangle_2020,tang_Tuning_2020}.  This not only allows ready identification of \moire excitons via the absorption spectra, but also suggests both of the \moire excitons may strongly couple with light and form stable \moire polaritons.

We first identify the existence of \moire lattice and \moire excitons before enclosing the hBL in a cavity. As shown by comparison of the hBL and ML reflection spectra in Fig.~\ref{fig:Schematic}c, the ML \MoSe A exciton is split into two \moire excitons in the hBL, X$_1$ and X$_2$, separated by about 40~meV, both with pronounced absorption, consistent with calculations (Fig.~\ref{fig:Schematic}b-c) and temperature dependence of the hBL (Extended Data Fig.2).

\begin{figure*}[t]
  \includegraphics[width=\linewidth]{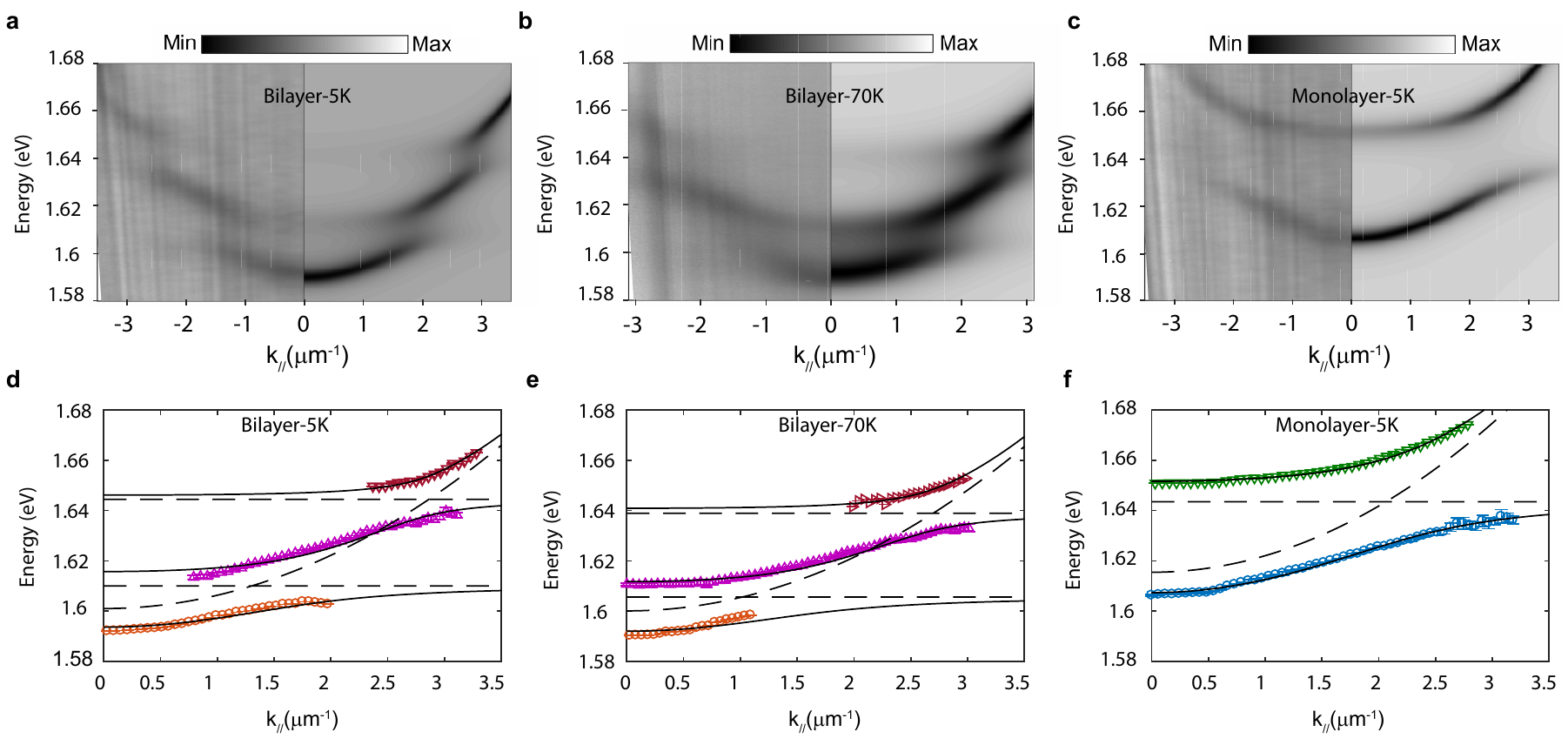}
	\caption{{\bfseries Strong coupling and dispersions of \moire and ML polaritonss.} $\bold{a-c}$, Angle resolved white light reflection spectra, demonstrating strong coupling between \moire exciton and cavity photon at 5K in $\bold{a}$ and 70K in $\bold{b}$, in comparison with strong coupling between ML exciton and cavity photon at 5K in $\bold{c}$. The left/right panels show the measured/simulated results, respectively. $\bold{d-f}$, The polariton energies vs. in-plane wavenumber k$_{//}$ obtained from $\bold{a-c}$, respectively. The solid lines are fits to the polariton dispersions with the coupled harmonic oscillator model. The dashed lines are the fitted energies of the uncoupled cavity photon and excitons. The error bars on the energy data correspond to the $95\%$ confidence interval of the Lorentzian fit.}%The extracted coupling strength for the lower and upper \moire excitons are $\sim10$ and 8 meV for both 5 and 70K. The coupling strength for monolayer exciton is 17.1 meV. }
	\label{fig:Dispersion}
\end{figure*}

%\section{\Moire polaritons}
When the hBL is placed at the anti-node of a $\lambda$/2 microcavity (Fig.~\ref{fig:Schematic}d, see Extended Data Fig.3 for more details of the cavity), in place of the \moire excitons or the bare cavity, three dispersive modes are observed up to 70~K, as shown in Fig.~\ref{fig:Dispersion}a-b. The modes anti-cross at the two \moire exciton resonances, showing clearly strong coupling of both \moire excitons (X$_1$ and X$_2$) with the cavity photon. Emission, temperature dependence, and time resolved studies are provided in the Extended Data Fig.~4-6.

The measured dispersions (Fig.~\ref{fig:Dispersion}d-e) are described very well by calculations based on a three coupled oscillator model with a Hamiltonian:
       \[
      H=\begin{bmatrix}
     E_{X_1} & 0 & \Omega_{1} \\
     0 & E_{X_2} & \Omega_{2} \\
     \Omega_{1} & \Omega_{2} & E_c
     \end{bmatrix}.
     \]
Here $E_c$ is the energy of the cavity mode, $E_{X_1, X_2}$ are the energies of the two \moire excitons, and $\Omega_{1,2}$ are their coupling strengths with the cavity photon. The three distinct dispersions measured correspond to the three new, light-matter hybrid eigen-modes of the Hamiltonian: the upper (UP), middle (MP) and lower (LP) polaritons.
The fitted $E_{X_1, X_2}$ agree with the independently measured \moire excitons energies (Fig.~\ref{fig:Schematic}c) within 5 meV, where the difference is commonly observed as a result of strain due to deposition of the top mirror.
The fitted $\Omega_{1,2} =10.1\pm0.3$~meV and $8.5\pm0.3$~meV at 4~K and change slightly to $9.6\pm0.4$~meV and $8.7\pm0.4$~meV at 70~K. From independent measurement, the exciton half linewidths increase from $\gamma_{X_1, X_2}=7.5$~meV and 8.4~meV at 4~K to 8.8~meV and 8.9 meV at 70~K (Extended Data Fig.5), which are mainly due to inhomogeneous broadening \footnote{The oscillator strengths of the \moire and ML excitons correspond to radiative linewidths of 100s~$\mu$eV and radiative decay rates of 100s~fs. Therefore the measured linewidth of both types of excitons are still dominated by inhomogeneous broadening. The measured photoluminescence decay times are a few picoseconds for both \moire and ML excitons (Extended Data Fig.6), which are consistent with the expected very short radiative lifetime and are lengthened due to energy relaxation dynamics. The \moire excitons have broader linewidths (half width of $7 - 8$ meV) than the ML excitons (about 2 meV), likely due to inhomogeneity in lattice alignment or strain distribution introduced during transfer and stacking of the two MLs.}.
The strong coupling condition $\Omega_{1,2}>(\gamma_{c}+\gamma_{X_{1,2}})/2$ is satisfied up to 70~K.

In comparison, a similar cavity enclosing a ML \MoSe also exhibits clearly strong coupling of the ML \MoSe A-exciton and the cavity photon (Fig.~\ref{fig:Dispersion}c and f). Fitting the dispersions of the polaritons yield $\Omega_{XA}$=17.1$\pm$0.1 meV, which is comparable to but slightly larger than $\sqrt{\Omega_{1}^2+\Omega_{2}^2}$=13.2 meV. This confirms that most of the oscillator strength of the ML \MoSe A-exciton is distributed to the \moire states X$_1$ and X$_2$. The reduction may be due to additional, higher-energy \moire states \cite{alexeev_Resonantly_2019a,tang_Simulation_2020} or additional disorder introduced into the bilayer during stacking of the MLs.

 \section{Zero-dimensionality of \moire excitons}
\begin{figure*}[t]
  \includegraphics[width=\linewidth]{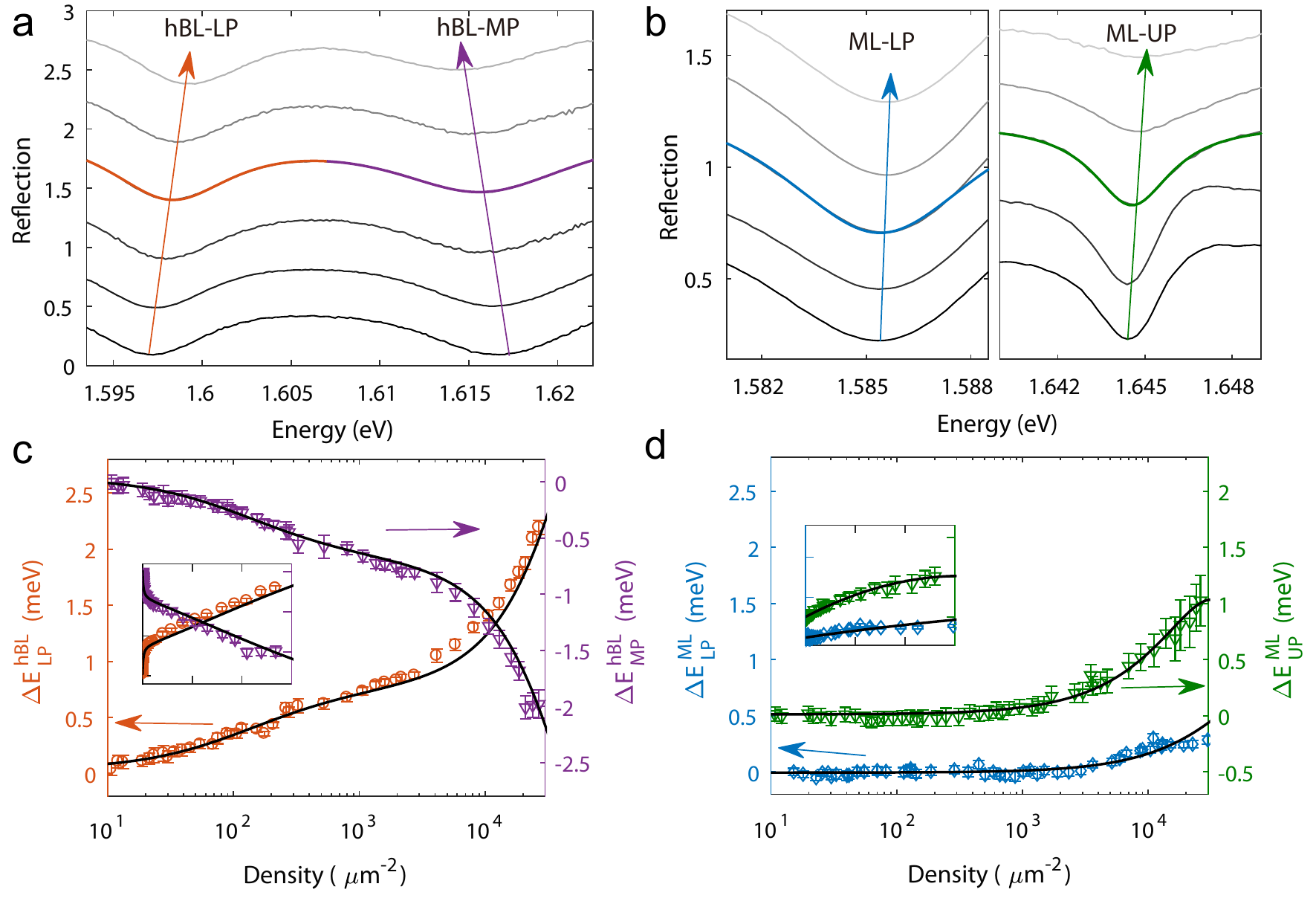}
	\caption{{\bfseries Nonlinearity of the \moire hBL and ML polaritons.} $\bold{a}$, Reflection spectra of \moire LP and MP at zero detuning for different pumping densities. From bottom to top, the carrier density increases from $17~\mu$m$^{-2}$ to $1.8\times 10^4~\mu$m$^{-2}$.
 $\bold{b}$, Reflection spectra of \MoSe ML LP (left) and UP (right). The cavity is red detuned from the exciton by 42.5~meV. From bottom to top, the carrier density increases from $22~\mu$m$^{-2}$ to $1.5\times 10^4~\mu$m$^{-2}$.
In $\bold{a,b}$, the colored solid lines are examples of Lorentzian fits. The arrows are guides for the eyes.
$\bold{c,d}$  Shift of polariton energies vs. the carrier density (log scale) obtained from $\bold{a}$ and $\bold{b}$, respectively. Insets show the density in the linear scale, plotted over the same ranges of the horizontal and vertical axes as the main figure. Solid lines are calculations using parameters obtained from fitting the data in Fig.~\ref{fig:nonlinearity} $\bold{a-c}$. The error bars on the energy data correspond to the $95\%$ confidence interval of the Lorentzian fit.
 }
    \label{fig:analysis}
\end{figure*}

With robust polariton modes formed with both \moire hBL and ML excitons, we study the effect of the underlying \moire lattice via their nonlinear response to the excitation density $n$.
We focus on $n<n_{Mott}=1/a_B^2\sim 10^{6}~\mu$m$^{-2}$, for $n_{Mott}$ the Mott density and $a_B\sim 1$~nm the Bohr radius \cite{chernikov_Population_2015}, and vary $n$ from $10$ $\mu$m$^{-2}$ up to $3\times10^{4}$ $\mu$m$^{-2}$ using a resonant 150 fs pulsed laser (see Methods for details) \cite{scuri_Large_2018a,emmanuele_Highly_2020,gu_enhanced_2019,kravtsov_Nonlinear_2019}.
A few representative reflectance spectra at different excitation densities are shown in  Fig.~\ref{fig:analysis}a-b for the hBL- and ML-cavities, respectively. Fitting the absorption dips in the spectra with Lorentzian functions, we obtain the polariton energies as plotted in Fig.~\ref{fig:analysis}c-d.

With increasing excitation densities, the \moire LP and MP shift symmetrically toward the \moire exciton resonance $E_{X_1}$ while their linewidths remain the same (Fig.~\ref{fig:analysis}a and Fig.~\ref{fig:analysis}c). These suggest reduced exciton-photon coupling strength, yet constant exciton energy or dephasing, which are typical properties of 0D excitons.
In sharp contrast, the ML LP and UP shift together to higher energies, accompanied by significant linewidth broadening. These suggest a much weaker saturation of the exciton-photon coupling strength but pronounced many-body effects, as expected of 2D excitons.

To analyze the density dependence quantitatively, we use the coupled-oscillator model to extract from the polariton spectra the three basic properties of excitons: the exciton energy $E_X(n)$, linewidth $\gamma_X(n)$, and photon coupling strength $\Omega(n)$ (see Methods for details). The results can be compared with well-established models for free 2D excitons:
\begin{align}
  E_{X}(n) & =  E_{X}(0)+\beta_{1}n-\beta_{2}n^2, \label{eq:EX}\\
\gamma_{X,n} & =  \gamma_{X}(0)+\alpha n, \label{eq:gamma}\\
 \Omega(n) & =  {\Omega(0)}/{\sqrt{1+\frac{n}{n_s}}} \label{eq:Omega}.
\end{align}
These equations describe the energy shift due to two and three-particle exchange interactions with coefficients $\beta_{1,2}$\cite{emmanuele_Highly_2020}, the linewidth broadening due to exciton-induced dephasing (EID) with coefficient $\alpha$ \cite{moody_Intrinsic_2015a}, and oscillator strength saturation due to Pauli blocking with a saturation density $n_s$ \cite{chernikov_Population_2015,Huang_Carrier_1990}.

\begin{figure*}[t]
  \includegraphics[width=.9\linewidth]{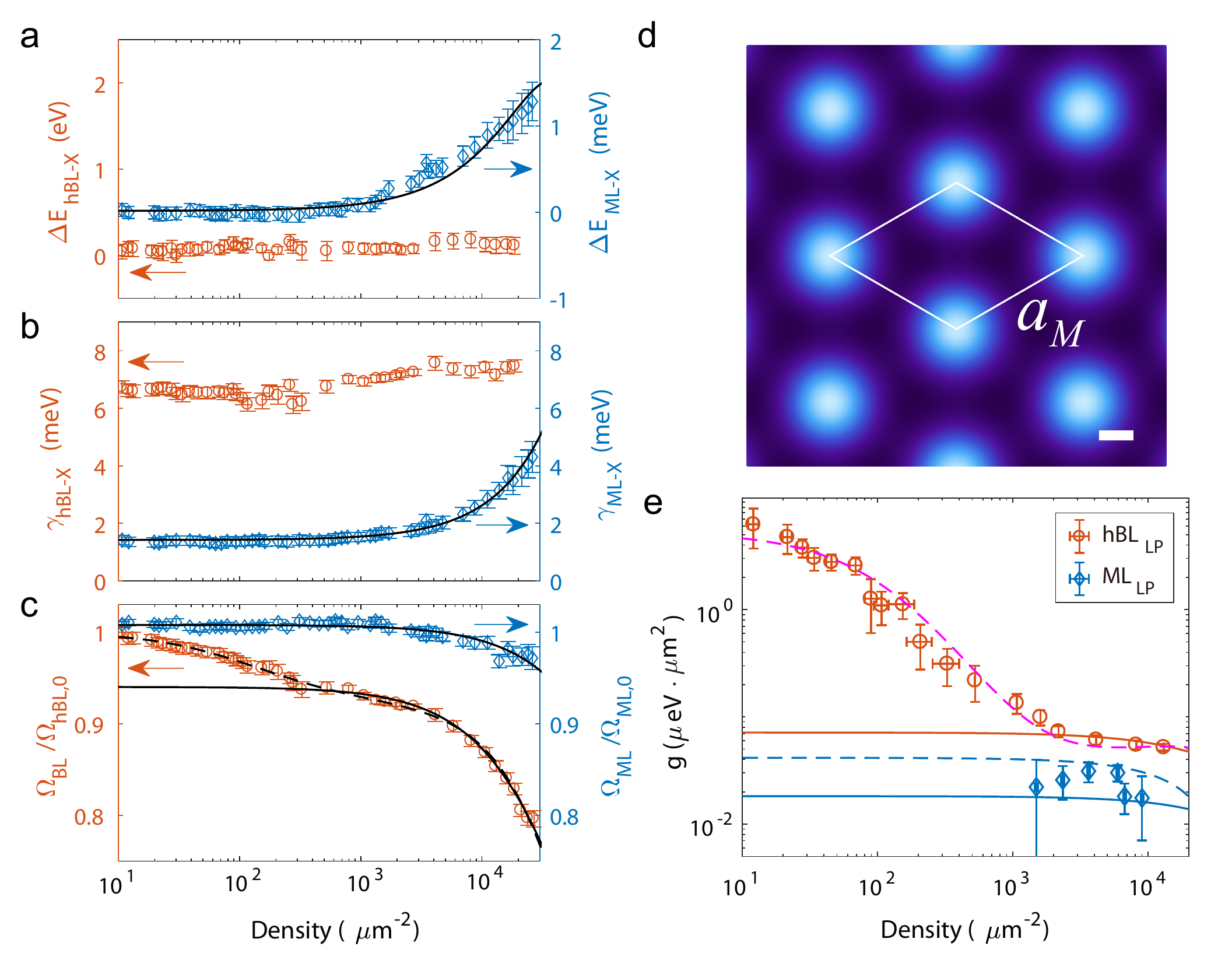}
	\caption{{\bfseries Enhanced nonlinearity by \moire lattice confinement.} $\bold{a-c}$, Shift of exciton energy $\Delta E$, half-linewidth $\gamma$, and normalized coupling strength $\Omega/\Omega_0$ of the hBL-LP (red) and ML-LP (blue) as a function of carrier density, obtained from the data in Fig.~\ref{fig:analysis}. The hBL $\Delta E_{hBL-X}$ and $\gamma_{hBL-X}$ (red circles in $\bold{a}$ and $\bold{b}$) are approximately constant. The ML $\Delta E_{ML-X}$ and $\gamma_{ML-X}$ (blue diamonds in $\bold{a}$ and $\bold{b}$) are fitted by a second order polynomial and a linear line, respectively (black solid lines in $\bold{a}$ and $\bold{b}$).
In $\bold{c}$, the black solid lines are fits with equation (3) with a constant $n_s$ for the \moire excitons at n$>$1000 $\mu$$m^{-2}$ and for ML excitons.
The black dashed line is a fit with a density-dependent effective saturation density $n_{eff,s}$, and is used for calculating the polariton energies in Fig.~\ref{fig:analysis} $\bold{c}$.
$\bold{d}$, Real-space distribution of the interlayer component of the X$_1$ state. The white lines mark a moir\'e unit cell. The scale bar is 1 nm.
$\bold{e}$, The measured (symbols) and fitted (lines) nonlinear coefficient $g$ vs. carrier density for the \moire hBL-LP (red) and ML-LP (blue).
%The solid lines with corresponding colors are calculations using the fitted polariton energies shown in Fig.~\ref{fig:analysis} $\bold{c,d}$.
The magenta dashed line and blue solid line are the calculations using the fitted polariton energies in Fig.~\ref{fig:analysis} $\bold{c}$ and $\bold{d}$.
%The blue dots are the calculation of $g_{ML-LP}$ considering a ML polariton system with zero exciton-photon detuning.
The hBL-LP is at zero detuing; the red solid line and magenta dashed lines correspond to fitted $\Omega_{hBL}$ using a constant and effective saturation density, $n_s$ and $n_{eff,s}$, in Eq~(3), respectively. For ML-LP, the blue solid and dashed lines correspond to the measured detuning and zero detuning, respectively. The error bars in $\bold{a}$-$\bold{c}$ and density in $\bold{e}$ are explained in Methods. The error bars of $g$ correspond to the $95\%$ confidence interval of the fit using $g(n)=\mid{dE(n)/dn}\mid$.  %The magenta dashed and blue solid lines also correspond to the fitted energy shift shown in Fig.~\ref{fig:analysis}$\bold{c}$ and $\bold{d}$, respectively.
%Since the nonlinearity depends on detuning, for the purpose of comparison, we also plot the expected nonlinearity of \moire and ML LPs both at zero-detuning (blue dots in Fig.~\ref{fig:nonlinearity}e).%cut-30
}
    \label{fig:nonlinearity}
\end{figure*}

Pronounced differences between \moire and ML-excitons are clearly seen in all three properties (Fig.~\ref{fig:nonlinearity}a-c). For the ML exciton, all three properties are described very well by Eqs.~\ref{eq:EX}-\ref{eq:Omega} for 2D excitons (blue diamonds and lines in Fig.~\ref{fig:nonlinearity}a-c). The exciton energy blueshifts by 1.5~meV due to exchange interactions, the linewidth broadens significantly by 3~meV due to EID. The fitted coefficients, $\beta_{1,ML}=(1.2\pm0.1)\times10^{-1}~\mu$eV$\cdot\mu$m$^{2}$, $\beta_{2,ML}=(2.9\pm0.4) \times 10^{-6}~\mu$eV$\cdot\mu$m$^{4}$, and $\alpha = 0.11\pm0.01~\mu$eV$\cdot\mu$m$^{2}$, all agree well with reported values \cite{emmanuele_Highly_2020, moody_Intrinsic_2015a}.
The coupling strength decreases slightly by up to $5\%$; the corresponding $n_{s,ML}=(2.8\pm0.4)\times 10^5~\mu$m$^{-2}$ is comparable to $1/a_B^2\sim 10^6~\mu$m$^{-2}$. These results confirm that the ML excitons behave as 2D particles; they also provide a consistency check of our density calibration.

In contrast to the ML excitons, the \moire excitons show no measurable energy shift, a much smaller line-broadening of $<1$~meV, and a much stronger saturation of the coupling strength by up to $20\%$ (Fig.~\ref{fig:nonlinearity}a-c, red circles). These can not be explained with the 2D exciton picture, but are expected of 0D excitons, as we discuss below.

In a \moire lattice, the exciton center-of-mass wavefunction is no longer a plane wave, but becomes localized near the potential minimum of each \moire cell, with a localization length $\ell$ ($a_B < \ell < a_M/2$, where $a_M$ is the moir\'e period), as illustrated in Fig.~\ref{fig:nonlinearity}d.
Evaluating $\ell$ for our device of $a_M=4.2$~nm yields $\ell =1.2$~nm.
Therefore the confinement leads to an increase of the effective local density $n_{eff}$ by $(a_M/2\ell)^2\sim 3$ at the potential minima, and thus enhanced exchange and dipole-dipole interactions. If the \moire exciton are 2D-like band excitons, the enhanced interactions should lead to correspondingly an enhanced energy shift and enhanced EID. On the contrary, we observe no energy shift and a much smaller line-broadening. Therefore, our observations show the \moire excitons in our device are no longer 2D band excitons.

On the other hand, the suppressed energy shift and linewidth broadening are characteristic of quantum dots with strong exciton blockade. Exciton blockade takes place when the interaction energy between two excitons in the same cell becomes greater than the exciton linewidth, so that multiple excitations correspond to distinct, quantized energy levels. In our hBL, the on-site exchange interaction $U_{exc, hBL}=\frac{1}{2\pi}(\frac{a_B}{w})^2 E_b \sim 40$~meV for binding energy $E_b\sim 250$~meV, and exciton wavefunction extension $w\sim$1~nm. The dipole interaction due to the inter-layer component is $U_{dd,hBL}\sim \frac{d}{a_B} U_{exc, hBL} \sim 30$~meV for dipole length $d\sim$0.7~nm \cite{yu_moire_2017}. These values agree with a more detailed calculation (Extended Data Fig.9) and are much larger than the exciton full linewidth of $2\gamma_{X_1}\sim 15$~meV. So we indeed expect exciton blockade in a \moire cell.  At the same time, since both the exchange and dipole-dipole interactions decrease quickly with distance, they are both suppressed for excitons in different \moire cells (Extended data Fig.9). The intra-cell exciton blockade, together with suppressed inter-cell interactions, lead to suppressed many-body effects for the single-exciton resonance, which manifests as absence of energy shift or EID, in agreement with our observations.

Consistent with exciton blockade, exciton-photon coupling saturates at one exciton per \moire cell, or, $n_s \sim 1/a_M^2 \sim (\pi a_B^2/a_M^2) n_{s, ML} \sim 6 \times 10^4 $, in excellent agreement with the fitted $n_{s,hBL}=(6.5\pm0.3)\times 10^4~\mu$m$^{-2}$ for $n\geq 1000$ $\mu$m$^{-2}$ (bottom solid line in Fig.~\ref{fig:nonlinearity}c). In contrast, if the \moire excitons are 2D band excitons, $n_s$ would have remained the same as $n_{s,ML}$, since the total oscillator strength is conserved in the band across the lattice. We note that the fit does not explain the abnormally strong saturation at very low densities of $n< 1000$ $\mu$m$^{-2}$, which we will discuss more later.

\section{\moire lattice-induced nonlinearity of \moire polaritons}
The distinctly different density dependence of the \moire polaritons shown above suggests the possibility of achieving a much higher polariton nonlinearity, even for polariton modes that are stable at high temperatures. Fig.~\ref{fig:nonlinearity}e shows the measured $g(n)=\mid{dE(n)/dn}\mid$ for both the \moire and ML LPs (symbols). %Since the nonlinearity depends on detuning, for the purpose of comparison, we also plot the expected nonlinearity of \moire and ML LPs both at zero-detuning (blue dots in Fig.~\ref{fig:nonlinearity}e).%cut-30

While the nonlinearity increases with decreasing density for both \moire and ML LPs, the \moire LPs show surprisingly a larger nonlinearity.
The nonlinearity of ML LPs originates primarily from exciton exchange-interactions and $g_{ML-LP}$ saturates below $n\sim 1000~\mu$m$^{-2}$ to $0.02~\mu$eV$\cdot\mu$m$^2$ ($0.04~\mu$eV$\cdot\mu$m$^2$) at the measured (zero) detuning, in agreement with reported values \cite{emmanuele_Highly_2020,kravtsov_Nonlinear_2019,tan_Interacting_2020}.
The nonlinearity of \moire LPs results primarily from exciton blockade. Based on the data at $n> 1000~\mu$m$^{-2}$, we obtain $g_{hBL-LP}$ about four times higher than $g_{ML-LP}$.

At very low densities of $n< 1000~\mu$m$^{-2}$, while the ML polaritons or excitons show no measurable shift, the \moire polaritons show clearly saturation-induced shifts down to $n\sim10~\mu$m$^{-2}$ (Fig.~\ref{fig:analysis}c), corresponding to a stronger exciton saturation than expected from exciton blockade, exciton-interactions, or effects of trions and defect states (Fig.~\ref{fig:nonlinearity}c)
\footnote{The strong saturation observed at very low excitation densities cannot be explained by trions or states in deeper trapping potentials. These states would have a small initial oscillator strength and lower resonance energy, so the increase of saturation density would have been accompanied by an increase of the coupling strength and exciton energy, in contradiction with our observations.}.
However, this abnormally large nonlinearity is repeatable over multiple measurements in multiple devices (see Extended Data Fig.7 for another example), suggesting hidden mechanisms for large polariton nonlinearity in \moire lattices.
Phenomenologically, $\Omega_{X1}(n)$ over the full density range can be described very well by Eq.~\ref{eq:Omega} if $n_s$ is replaced by a density-dependent effective saturation density $n_{eff,s}=\tilde{n}_{s}(1-A e^{-n/B})$, for fitted $\tilde{n}_{s}=(4.4\pm0.1)\times10^4~\mu$m$^{-2}$, $A = 0.98\pm0.04$ and $B=(7.4\pm 0.5)~\mu$m$^{-2}$ (Fig.~\ref{fig:nonlinearity}c). %$\tilde{n}_s$ corresponds to saturation density in the limit of very high excitation densities; it  agrees with $n_s$ obtained from fitting the high density data only as well as the estimate based on $1/a_M^2$. The density $\tilde{n}_s(1-A) =(8.8\pm0.4)\times10^2~\mu$m$^{-2}$ corresponds to saturation in the limit of $n\rightarrow 0$, and is reduced by 50 times compared to $\tilde{n}_s$ in our sample.

The polariton nonlinearity is a key figure of merit that distinguishes polariton systems from pure photon systems. Together with the robust coherence enforced by the photon component, it gives rise to
%many remarkable phenomena, such as polariton parametric amplification\cite{saba_high-temperature_2001}, polariton condensation\cite{deng_Condensation_2002}, spontaneous limit-cycle oscillations\cite{kim_Emergence_2020} and ultrafast optical switching\cite{liew_optical_2008,gao_polariton_2012,sturm_Alloptical_2014, ballarini_Alloptical_2013,dreismann_Subfemtojoule_2016}. Strong nonlinearity is also crucial for polariton blockades \cite{delteil_Polariton_2019,munoz-matutano_Emergence_2019}, quantum simulation, and
a wide range of novel nonlinear many-body and quantum phenomena \cite{ballarini_Alloptical_2013, walker_Ultralowpower_2015a, berloff_realizing_2017, delteil_Polariton_2019,munoz-matutano_Emergence_2019,kim_Emergence_2020}.
%,sturm_Alloptical_2014, dreismann_Subfemtojoule_2016,

For polaritons made of 2D excitons, however, larger $g$ is obtained only at the compromise of $\Omega$, or the stability of polariton modes. This is because the exciton-exchange interaction and the exciton-photon coupling strength saturation, the two main contributions to $g$, both decrease with the exciton reduced mass $\mu$: $g_{exc} \sim E_b a_B^2 \propto 1/\mu$ and $g_{sat}\propto a_B^2 \propto 1/\mu^2$. Yet the exciton-photon coupling strength $\Omega \propto \sqrt{f} \propto 1/a_B \propto \mu$. The highest polariton nonlinearity is achieved in single or few quantum-well GaAs microcavities \cite{delteil_Polariton_2019,munoz-matutano_Emergence_2019},
%,walker_Ultralowpower_2015a,ferrier_Interactions_2011a
with $g\sim 3~\mu$eV$\cdot\mu$m$^2$ and $\Omega$ only $\sim$3~meV.
Wide-bandgap semiconductors, organic crystals and ML TMDCs all feature greater $\Omega$ and high-temperature polaritons, yet a much weaker polariton nonlinearity \cite{daskalakis_Nonlinear_2014a,scuri_Large_2018a,barachati_Interacting_2018,kravtsov_Nonlinear_2019}.
Higher order excitations in TMDCs, such as trions with strong band-filling effect \cite{emmanuele_Highly_2020,tan_Interacting_2020} and 2s excitons with a larger Bohr radius \cite{gu_enhanced_2019}, have shown enhanced nonlinearity, but again at the compromise of stability.

The \moire polariton system uniquely combines strong nonlinearity, due to quantum confinement of excitons within each \moire cell, and a large total photon coupling strength, due to collective coupling among the cells. It thereby provides a new route to achieve strong nonlinearities simultaneously with robust exciton-photon coupling.

\section{Conclusion}
In summary, we have demonstrated the first polariton system formed via collective coupling of a 2D lattice of 0D excitons with light in a microcavity.
%The 0D nature of the excitons is revealed by distinctly different density dependence of the polariton and exciton properties compared to polaritons made of 2D excitons, including suppressed exciton- energy shift, suppressed excitation-induced dephasing, and strong saturation-induced polariton nonlinearity due to exciton blockade.
The system therefore introduces quantum-dot like nonlinearity into cooperatively coupled solid state system, opening a door to novel quantum many-body physics and polariton devices \cite{ritsch_cold_2013,gonzalez-tudela_subwavelength_2015,leonard_Supersolid_2017,yu_Electrically_2020}.
%Unlike single quantum dot systems, the close-packing and collective coupling of the 0D exciton in a \moire lattice lead to a large exciton-photon coupling strength and robust polariton modes up to 70 K, opening a door to novel quantum many-body physics and polariton devices \cite{yu_Electrically_2020}.
Polariton blockade and a strongly-correlated polariton gases may become possible with reduced inhomogeneous broadening of \moire excitons, improved cavity quality factors, and a better understanding of the abnormal enhancement of the polariton nonlinearity at very low excitation densities.
Electrical gating and electrical field tuning of the heterobilayer can be implemented to further control the nonlinearity and many-body phenomena in the \moire polariton system.% \cite{tang_Tuning_2020}.

\section*{Methods}

\noindent\textbf{Sample fabrication.}
 %The WS2/MoSe2 heterobilayers are fabricated using a dry-transfer method with a polyethylene terephthalate (PET) stamp.
The ML \MoSe, \WS and few layer hBN flakes were obtained by mechanical exfoliation from bulk crystals. A polyethylene terephthalate (PET) stamp was used to pick up the top hBN, \WS ML, \MoSe ML, and the bottom hBN under microscope. The \WS and \MoSe monolayer MLs were rotationally aligned to about $0^\circ$ in the heterobilayer.

The bottom half of the cavity consists of 16 pairs of SiO$_2$/TiO$_2$ DBR with a $\lambda/4$ SiO$_2$ layer. The heterostructure on the PET stamp was stamped onto the bottom half of the cavity and the PET was dissolved in dichloromethane for six hours at room temperature. Then 67 nm PMMA film was spin coated on the top of heterobilayer to form the second half of the $\lambda$/2 cavity. Then a silver film of 40 nm thick was deposited using electron beam evaporation as the top mirror of the cavity.

\noindent\textbf{Optical measurements.}
For low temperature measurements, the sample was kept in a 4~K cryostat (Montana Instrument).  The excitation and collection were carried out with a home-built confocal microscope with an objective lens with a numerical aperture (NA) of 0.42. To characterize the dispersion of the polariton device, we performed angle-resolved reflection measurement using white light from a tungsten halogen lamp. The white light was focused on the sample with a beam size of 10 $\mu$m in diameter. To perform power dependent reflection measurements, we used a 150 fs-pulse laser as the light source with a repetition rate of 80 MHz and focused beam size on the sample of around 1.5 $\mu$m in diameter. For the hBL, the laser was centered between LP and MP to simultaneously measure both modes, while the intensity at UP is negligible (Extended Data Fig.8 for an example spectrum of the laser). For the ML, due to the larger energy separation of the LP and UP, they were measured separately with the laser centered at each, respectively. The signal was detected using a Princeton Instruments spectrometer with a cooled charge-coupled camera.% The laser is resonant with the lower exciton $X_1$ for \moire polariton measurements and resonant with the LP and UP modes, respectively, for the ML measurement.

\noindent\textbf{Theory of \moire excitons} The Hamiltonian for the \moire excitons is
\begin{equation}
\mathcal{H} =\begin{pmatrix}
 E_G + \frac{\hbar^2 \boldsymbol{k}^2}{2 M_\text{X} }  &
 w(1 + e^{-i \boldsymbol{b}_+ \cdot \boldsymbol{r} } + e^{-i \boldsymbol{b}_- \cdot \boldsymbol{r}})
 \\
 w(1 + e^{i \boldsymbol{b}_+ \cdot \boldsymbol{r} } + e^{i \boldsymbol{b}_- \cdot \boldsymbol{r}})                                              & E_G + \delta_0+ \frac{\hbar^2 (\boldsymbol{k}-\boldsymbol{\kappa})^2}{2 M_\text{IX} }
\end{pmatrix}
\end{equation}
where $E_G$ is an energy constant, $M_\text{X}$ and $M_\text{IX}$ are respectively the effective masses for intralayer
and interlayer excitons, $ \boldsymbol{\kappa}= (0, 4\pi/(3a_M))$ accounts for the momentum mismatch between the two
excitons, and $\delta_0$ is an energy offset. The off-diagonal terms are derived from interlayer tunneling,
and $\boldsymbol{b}_{\pm} = 4\pi/(\sqrt{3} a_M)(\pm 1/2,\sqrt{3}/2)$. The moir\'e period is $a_M=a_0/\sqrt{\delta^2+\theta^2}$, where $a_0=(a_{\text{MoSe}_2}+a_{\text{WS}_2})/2= 3.26$ \AA  and $\delta=|a_{\text{MoSe}_2}-a_{\text{WS}_2}|/a_0 \approx 4$\%.
We use the following parameter values, $E_G=1.61$eV, $\delta_0=-20$ meV,$w=14$ meV,
$M_\text{X}=0.82 m_0$, $M_\text{IX}=0.71 m_0$, where $m_0$ is the electron bare mass. The energy spectrum of the \moire excitons is obtained by diagonalizing the moir\'e
Hamiltonian $\mathcal{H}$ using plane-wave expansion, and is shown in Fig.~\ref{fig:Schematic}b. For the lowest-energy
exciton X$_1$, its interlayer component is plotted in Fig.~\ref{fig:nonlinearity}d, which shows spatial localization.

\noindent\textbf{Polariton density calibration}
  %The density dependence of polariton is measured using a pulse laser with repetition rate of f=80 MHz, and pulse width of 150 fs. The reflected light is spectrally resolved to obtain the absorbed intensity
%\begin{align}
%\label{eq:Absorption}
%   I_{absorb} = I_{incident}(\omega)-I_{reflect}(\omega),
%\end{align}
%so the absorption is $\alpha=1-\frac{I_{reflect}(\omega)}{I_{incident}(\omega)}$. The injected polariton density is
%\begin{align}
%\label{eq:Density}
%   n_{LP,UP} = \frac{P_{incident}(\omega)\times \alpha}{f\times E_{LP,MP}\times A},
%\end{align}
%P is the incident power at the frequency of polariton resonance,$E_{LP,MP}$ is the energy of lower or medium polaritons, and A=1.5 $\mu$m$^2$ is the beam area.
In this section, we will use the bilayer device data as an example;
the same procedure applies to the monolayer device.
To extract the polariton density:
First, we fit the reflection spectra in Fig.~\ref{fig:analysis} using :
\begin{align}
\label{eq:R}
   R = 1-Absor_{LP}-Absor_{MP},
\end{align}
where $Absor_{LP}$ and $Absor_{MP}$ represent absorption by the lower polariton ($LP$) and middle polariton ($MP$) in the bilayer, respectively, and they are described by Lorentzian functions:
\begin{align}
\label{eq:Absor}
   Absor_{LP,MP}(E) = \frac{A_{LP,MP}}{(E-E_{LP,MP})^2+\gamma_{LP,MP}^2},
   \end{align}

Resonance energy $E_{LP,MP}$, linewidth (HWHM) $\gamma_{LP,MP}$, and absorption amplitude $A_{LP,MP}$ are fitting parameters with uncertainties $\delta_{E_{LP,MP}}, \delta_{\gamma_{LP,MP}}$ and $\delta_{A_{LP,MP}}$ corresponding to $95\%$ confidence intervals.

  We use pulsed laser with pulse duration of 150 fs and repetition rate $f$ of 80 MHz. We calculate the polariton density $n$ injected per pulse.  The average power of the laser is $P$. The laser profile can be fitted with Gaussian function:
  \begin{align}
\label{eq:Gaussian}
   G(E)=\frac{A_{laser}}{\gamma_{laser}\sqrt{\pi/2}}e^{-\frac{2(E-E_{laser})^2}{\gamma_{laser}^2}}
\end{align}
  where $A_{laser}$ is the area of the Gaussian function, and $\gamma_{laser}$ is the linewidth. The power absorbed by LP and MP,  $P_{LP,MP}$ can be calculated using the convolution:
 \begin{align}
\label{eq:Power}
   P_{LP,MP}=\frac{P\int G(E)Absor_{LP,MP}}{A_{laser}}=\frac{P\int G(E)\frac{A_{LP,MP}}{4(E-E_{LP,MP})^2+\gamma^2_{LP,MP}}}{A_{laser}}
   \end{align}
  The total carrier density including both LP and MP created per pulse and its error $\delta_n$ can be calculated by:
   \begin{align}
\label{eq:density-method}
n=(n_{LP}+n_{MP})/A_{beam}=(\frac{P_{LP}}{fE_{LP}}+\frac{P_{MP}}{fE_{MP}})/A_{beam}.\\
     \delta_n=\sqrt{\delta^2_{n_{LP}}+\delta^2_{n_{MP}}}/A_{beam}
\end{align}
  Where, A$_{beam}$=1.5 $\mu$m, is the beam area, and $\delta_{n_{LP,MP}}=n_{LP,MP}\sqrt{(\frac{\delta_{P_{LP,MP}}}{P_{LP,MP}})^2+(\frac{\delta_{E_{LP,MP}}}{E_{LP,MP}})^2}$

  \noindent\textbf{Extraction of exciton energy, linewidth, and coupling strength at different densities}
To analyze the results quantitatively, we extract the density dependence of the exciton properties from the measured polariton spectra. In the heterobilayer cavity, we focus on the MP and LP modes and neglect changes caused by the X$_2$, since X$_2$ is at a much higher energy, and its change only negligibly affects MP and LP (see Extended Fig.10 for details). The cavity resonance energy $E_c$ and linewidth $\gamma_c$ are assumed to change negligibly with exciton density. Using the two coupled oscillator mode, the energies of the MP and LP of the heterobilayer cavity, as well as of the UP and LP of the ML-cavity can be extracted. In the following, we will use the bilayer device data as an example; the same procedure applies to the monolayer device.
\begin{align}
\label{eq:E_pol}
   E_{\textrm{LP,MP}}(n) = & \frac{1}{2}\bigg[ E_{X}(n)+E_{C}+i( \gamma_{C}+\gamma_{X}(n) ) \bigg] \nonumber\\
    \pm&\sqrt{\Omega(n)^2+\frac{1}{4}\bigg[E_{X}(n)-E_{C}+i(\gamma_{C}-\gamma_{X}(n))\bigg]^2}
\end{align}
Here the cavity resonance, $E_C$, and cavity half-linewidth $\gamma_{C}$, do not change with carrier density. Therefore, from the measured polariton energies and half-linewidth, $\gamma_{LP, MP}$, we obtain the density dependence of exciton energy, $E_{X}$, half-linewidth, $\gamma_{X}$, and exciton-photon coupling strength, $\Omega$:

Exciton energy $E_{X}$ and its uncertainty $\delta_{E_X}$ :
\begin{align}
\label{eq:hBL-X}
   E_{X}(n)=E_{LP}(n)+E_{MP}(n)-E_{C}\\
   \delta_{E_X}=\sqrt{\delta_{E_{LP}}^2+\delta_{E_{MP}}^2+\delta_{E_C}^2}
\end{align}
where, $E_C$ and $\delta_{E_C}$ are the cavity resonance and its uncertainty obtained by fitting the angle resolved reflection spectrum in Fig.~\ref{fig:analysis}, and they do not change with polariton densities.

Exciton linewidth $\gamma_{X}$ and its uncertainty $\delta_{\gamma_{X}}$:
\begin{align}
\label{eq:hBL-Linewidth}
   \gamma_{X}(n)=\gamma_{LP}(n)+\gamma_{MP}(n)-\gamma_{C}\\
   \delta_{\gamma_X}=\sqrt{\delta_{\gamma_{LP}}^2+\delta_{\gamma_{MP}}^2+\delta_{\gamma_C}^2}
\end{align}
where, $\gamma_C$ and $\delta_{\gamma_C}$ are the cavity linewidth and uncertainty obtained by fitting the reflection spectrum of the bare cavity in Fig.~\ref{fig:Schematic}d, and they do not change with polariton densities.

Coupling strength $\Omega$:
\begin{align}
\label{eq:hBL-Couplingstrength}
   \Omega(n)=\frac{1}{2}\sqrt{\left[E_{LP}(n)-E_{MP}(n)+i(\gamma_{LP}(n)-\gamma_{MP}(n)) \right]^2-\left[E_{X}(n)-E_{C}+i(\gamma_{X}(n)-\gamma_{C}) \right]^2}
   %\delta_{\Omega}=\frac{1}{2}\sqrt{\left|E_{LP}-E_{MP}+i(\gamma_{LP}-\gamma_{MP}) \right|^2-\left|E_{X}-E_{C}+i(\gamma_{X}-\gamma_{C}) \right|^2}
\end{align}

\noindent\textbf{Dipole-dipole interaction}
The interlayer component of \moire excitons contributes to dipole-dipole interactions. The wave function for interlayer component of a \moire exciton localized near the potential minimum around the origin can be described as
\begin{equation}
    W(\bm{r}_1,\bm{r}_2) = x_{\text{IX}} \Phi(\frac{\bm{r}_1+\bm{r}_2}{2}) \phi(\bm{r}_1-\bm{r}_2),
\end{equation}
where $x_{\text{IX}}$ is the interlayer component weight, $\Phi$ and $\phi$ are respectively the center-of-mass and relative-mass wave function. We approximate $\Phi$ by a Gaussian
$\Phi(\boldsymbol{R}) = \frac{1}{\sqrt{\pi} \ell} e^{-R^2/(2 \ell^2)}$, and $\phi(\boldsymbol{r})=\sqrt{8/\pi}(1/a_B) e^{-2r/a_B} $, where $\boldsymbol{R}=\frac{\bm{r}_1+\bm{r}_2}{2}$, $\boldsymbol{r}=\bm{r}_1-\bm{r}_2$, $\ell$ is the localization length that can be estimated using the exciton state shown in Fig.~\ref{fig:nonlinearity}d, and $a_B$ is the Bohr radius.

The dipole-dipole interaction between two excitons localized at different moir\'e sites can be approximated as
\begin{equation}
\begin{aligned}
U_{dd} (\boldsymbol{L}) = & \int d \bm{r}_1 \int d \bm{r}_2 \int d \bm{r}_3                            \int d \bm{r}_4 |W(\bm{r}_1,\bm{r}_2)|^2
|W(\bm{r}_3-\boldsymbol{L},\bm{r}_4-\boldsymbol{L})|^2 \\                   &[V_S(\bm{r}_1-\bm{r}_3)+V_S(\bm{r}_2-\bm{r}_4)-V_D(\bm{r}_1-\bm{r}_4)-V_D(\bm{r}_2-\bm{r}_3)]\\
= &  2 x_{\text{IX}}^2 \int \frac{d^2 \bm{q}}{(2\pi)^2} V(\bm{q}) \frac{\exp(-q^2 \ell^2/2)}{[1+\frac{1}{64} a_B^{*2} q^2]^3} e^{-i \bm{q} \cdot \bm{L}},
\end{aligned}
\end{equation}
where $\bm{L}$ is a moir\'e lattice vector, $V_S(\bm{r})=e^2/(\epsilon r)$ is the intralayer Coulomb interaction, $V_D(\bm{r})=e^2/(\epsilon \sqrt{r^2+d^2})$ is the interlayer Coulomb interaction with $d$ the interlayer distance, and $V(\bm{q})=\frac{2\pi e^2}{\epsilon q} (1-e^{-q d})$. To calculate $U_{dd} (\boldsymbol{L})$, we take $x_{\text{IX}} \approx 1/\sqrt{2}$, $d=0.65$ nm, $\epsilon = 5$, and $a_B \approx 1$ nm. Extended Fig.9 shows the onsite and nearest-neighbor dipole-dipole interactions as a function of twist angle. The onsite repulsion is sizable ($\sim$ 24 meV for a twist angle of 56.5$^\circ$), while the offsite repulsion is negligible.

\noindent\textbf{Data availability} Data are available on request from the authors.

\noindent\textbf{Competing interests} The authors declare that they have no competing financial interests.

\noindent\textbf{Author Contributions}
  H.D., L.Z.conceived the experiment. L.Z. performed the measurements.
  F. W. provided theoretical inputs. L.Z. and Z.Z fabricated the device. L.Z. and H.D. performed data analysis. S.H deposited the silver mirror. Y.C grew the bottom DBR mirror. K.W. and T.T grew hBN single crystals. H.D. and S.F. supervised the projects. L.Z and H.D. wrote the paper with inputs from other authors. All authors discussed the results, data analysis and the paper.
%\noindent\textbf{Competing financial interests:} The authors declare no competing financial interests.

\noindent\textbf{Acknowledgment}
We are grateful for helpful discussions with Duncan Steel and Mack Kira.
L. Z., S. H, S. F.and H. D.acknowledge the support by the Army Research Office under Awards W911NF-17-1-0312. L.Z. and H.D. also acknowledge the support by the Air Force Office of Scientific Research under Awards FA2386-18-1-4086 and by the National Science Foundation under Awards DMR-1838412. S. F. also acknowledges support from  the U.S. Department of Energy, Office of Science, Office of Basic Energy Sciences, under Awards DE-SC0017971. F. W. is supported by the Laboratory for Physical Sciences.  Y. C. acknowledge the support by the Young Scholar Fellowship Program by the Ministry of Science and Technology (MOST) in Taiwan, under Grant MOST 108-2636-M-006 -010.
K. W. and T. T. acknowledge support from the Elemental Strategy Initiative
conducted by the MEXT, Japan ,Grant Number JPMXP0112101001,  JSPS
KAKENHI Grant Numbers JP20H00354 and the CREST(JPMJCR15F3), JST.
%\noindent\textbf{Disclaimer} This report was prepared as an account of work sponsored agencies of the United States Government. Neither the United States Government nor any agency thereof, nor any of their employees, makes any warranty, express or implied, or assumes any legal liability or responsibility for the accuracy, completeness, or usefulness of any information, apparatus, product, or process disclosed, or represents that its use would not infringe privately owned rights. Reference herein to any specific commercial product, process, or service by trade name, trademark, manufacturer, or otherwise does not necessarily constitute or imply its endorsement, recommendation, or favoring by the United States Government or any agency thereof. The views and opinions of authors expressed herein do not necessarily state or reflect those of the United States Government or any agency thereof.
% \item[Competing interests] The authors declare that they have no competing financial interests.

\section*{References}
\bibliographystyle{naturemag}
\bibliography{Reference}%,SIReference}

%\newpage

\section*{}

%\begin{figure*}[t]
%  \includegraphics[width=\linewidth]{Figure3.png}
%	\caption{Tuning of interlayer coupling by twist angle. (a)RC spectrums at different twist angles. The detuning $\delta$ is the energy difference between interlayer exciton and intralayer exciton: $\delta=E_{inter}-E_{intra}$. The detunings are extracted by analyzing the RC spectrums.(b) schematics of exciton bands at different twist angles with intralyer exciton bands in red and interlayer exciton bands in blue. The momentum conserved interlayer coupling is achieved by band folding induced by moir\'e superlattice. (c) Ratio of oscillator strength from the two hybrid excitons $LHX_{MA}$ and $UHX_{MA}$, detuning, and coupling strength as a function of twist angles. The gray solid lines in the middle panel of c are quadratic fits to the angle dependence of detuning.
%        }
%	\label{fig:twist}
%\end{figure*}

\section*{}%Extended Data}
\newpage
%\noindent\textbf{Extended data figures and tables}
\renewcommand{\figurename}{Extended Data Fig.}
\setcounter{figure}{0}
%\section*{Figures}
\begin{figure*}[h]
	\includegraphics[width=\linewidth]{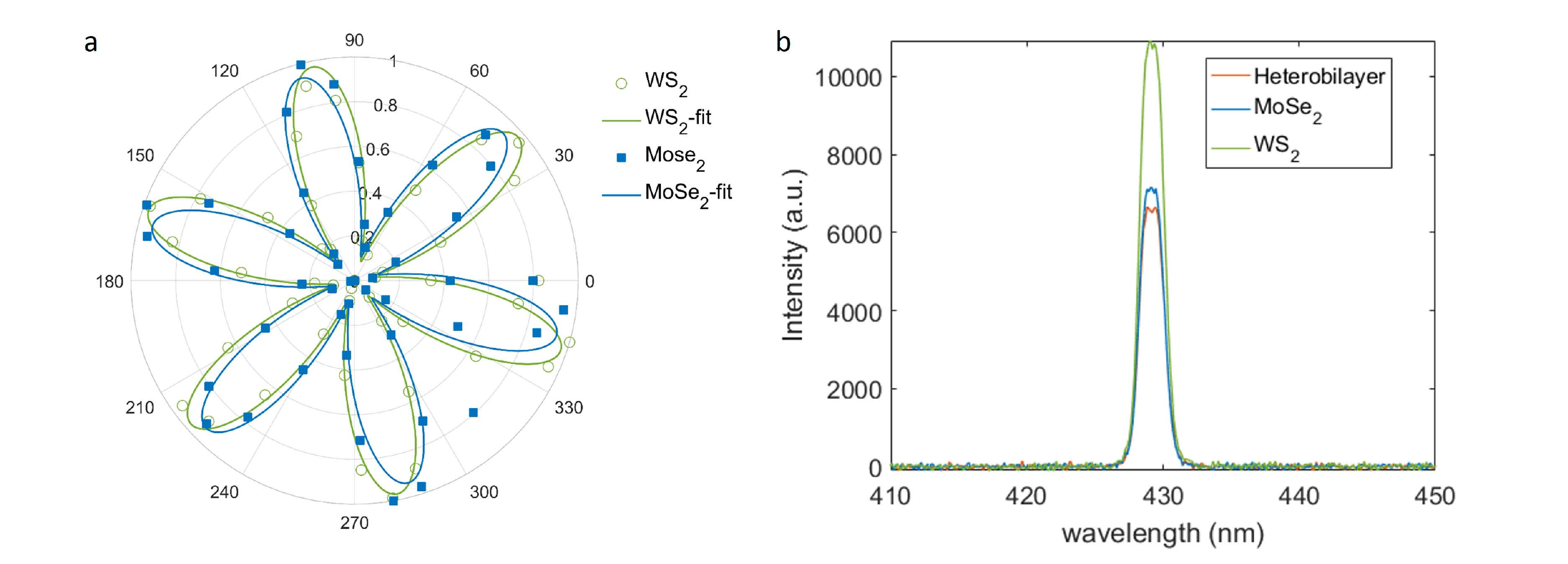}
	\caption{{\bfseries Heterobilayer twist angle.} Twist angle of hBL in the main text measured by second harmonic generation (SHG) spectroscopy. $\bold{a}$ The polarization-dependent SHG signal  measured on the ML \WS (green open circles) and MoSe2 (blue filling squares) regions of the hBL, and the corresponding fits with the sinusoidal functions (green and blue solid lines). $\bold{b}$, the SHG signal from ML WS2 , ML MoSe2 , and hBL regions, measured with the same experimental configurations. The suppressed SHG signal from hBL as a result of destructive interference indicates the stacking order is H-stacking. The twist angle is determined to be $56.5^\circ\pm0.8^\circ$ }
    \label{fig:twist}
\end{figure*}
\begin{figure*}[h]
	\includegraphics[width=\linewidth]{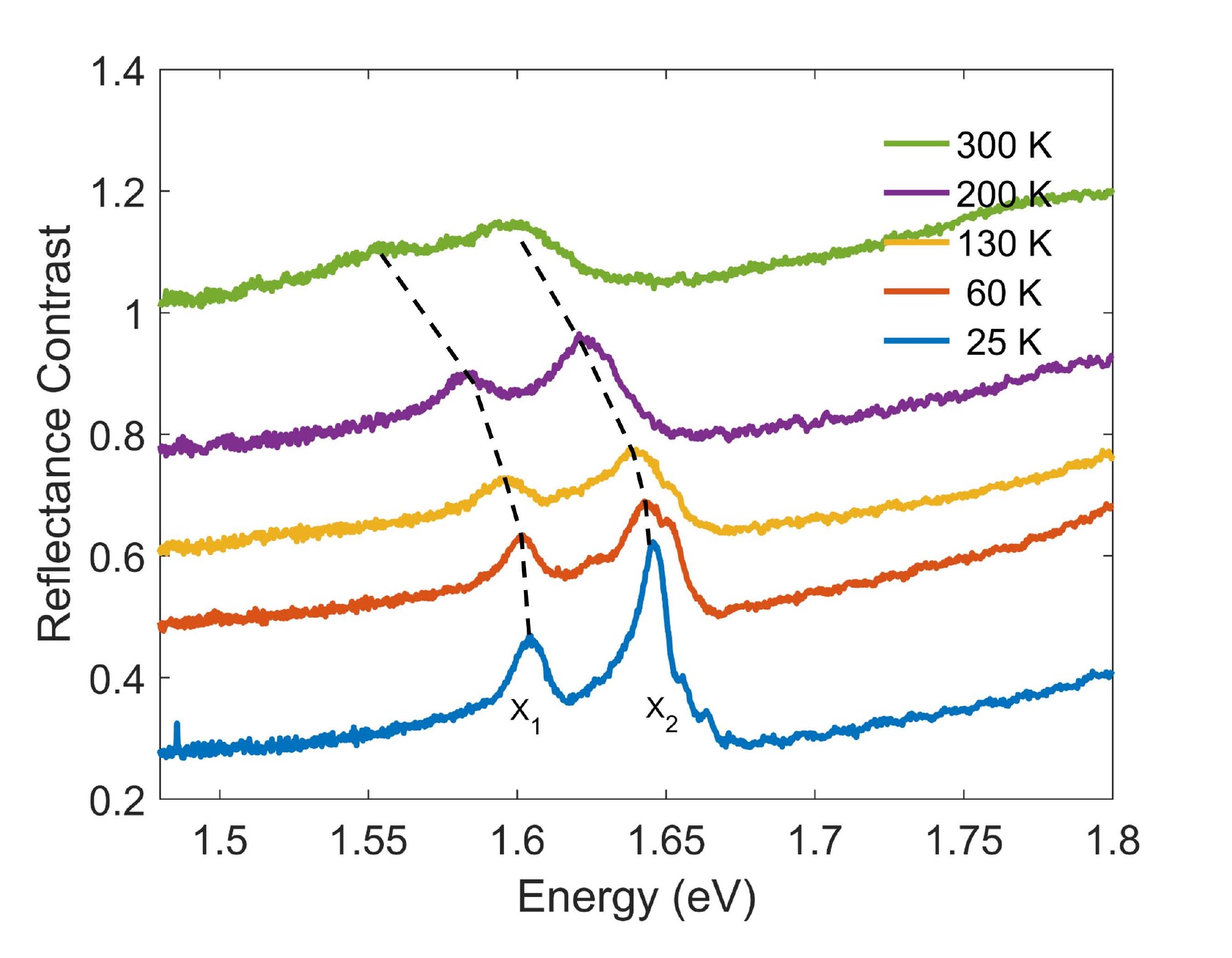}
	\caption{{\bfseries Temperature dependence of \moire exciton.} Temperature dependence of the \moire exciton X$_1$ and X$_2$ measured from a separate heterobilayer prepared on sapphire substrate. The black dashed lines are guides for eyes. The two exciton states can be well resolved up to 200K, which can exclude the possibility of charged exciton or trapped exciton by defect.}
    \label{fig:sample}
\end{figure*}
\begin{figure*}[t]
	\includegraphics[width=\linewidth]{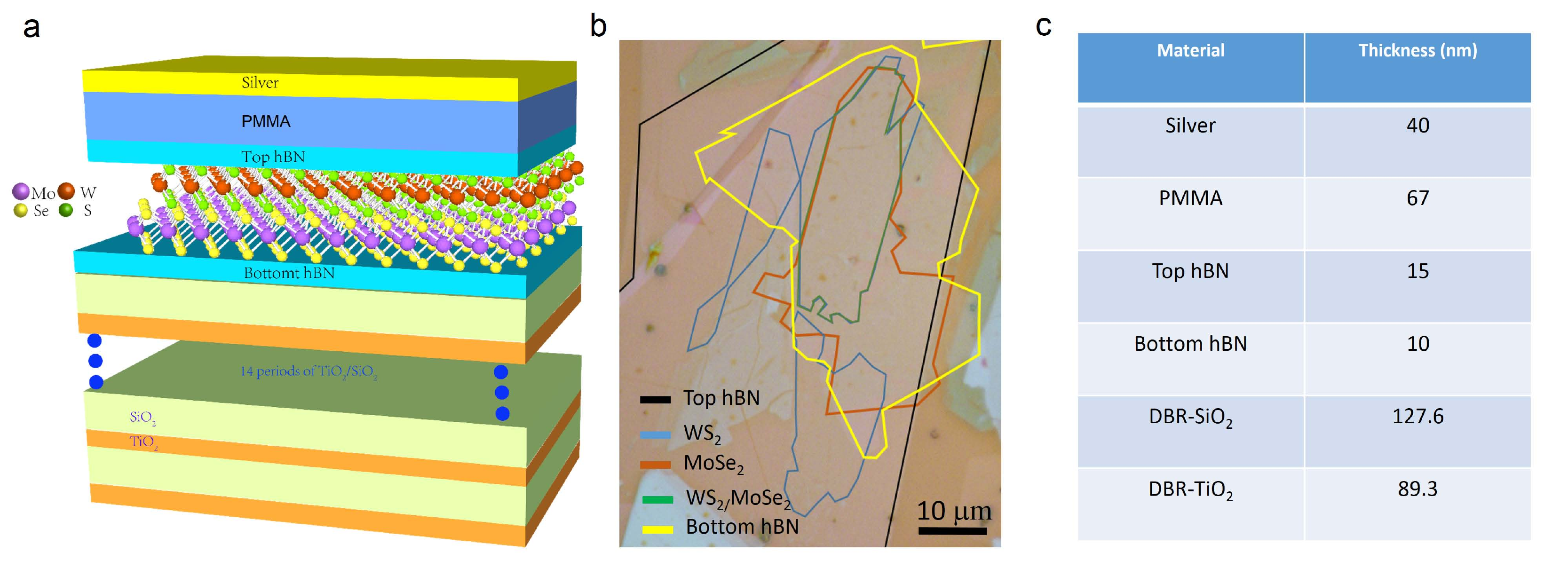}
	\caption{{\bfseries Schematic of the device.} $\bold{a}$ Schematic of the device shows the different layers of the heterostructure embedded inside a microcavity that consists of a bottom DBR and a top silver mirror. $\bold{b}$ Microscope image of the hBL on the top of DBR mirror, taken before depositing the PMMA layer and the silver mirror. $\bold{c}$ Thickness of each layer for the device.}
    \label{fig:sample}
\end{figure*}
\begin{figure*}[t]
	\includegraphics[width=\linewidth]{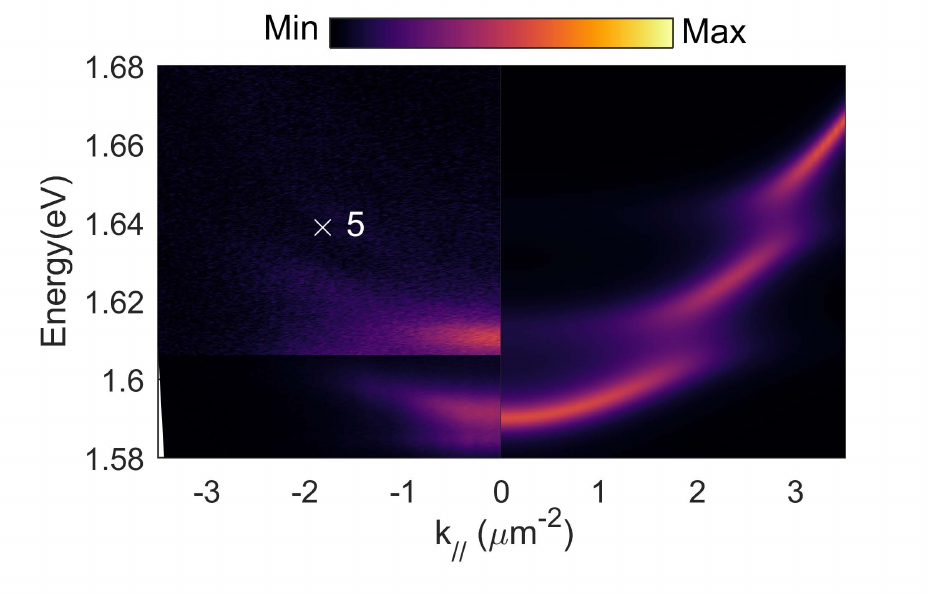}
	\caption{{\bfseries Photoluminescence from the \moire polariton.} Left panel: Angle resolved photoluminescence spectrum of hBL in cavity, excited by a continuous wave laser at the energy of 2.3 eV and power of 50 $\mu$W.  To enhance the visibility of states at higher energy, emission intensity above 1.607 eV is magnified by 5 times. Right panel: simulated angle resolved absorption, which agrees well with the measurement.}
    \label{fig:T-depen}
\end{figure*}

\begin{figure*}[t]
	\includegraphics[width=\linewidth]{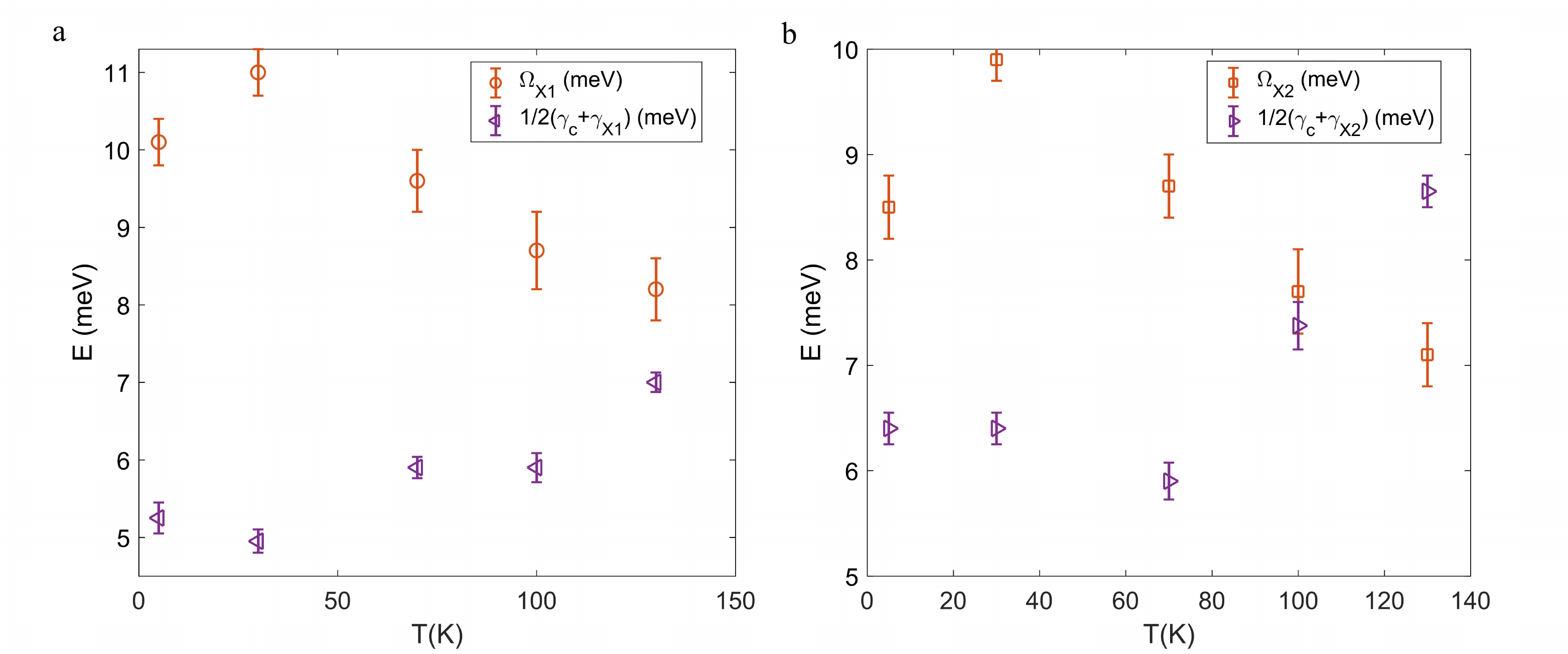}
	\caption{{\bfseries Transition from strong coupling to weak coupling driven by thermal broadening.} Strong coupling to weak-coupling transition measured by temperature dependence of $\Omega_{1}$ (red open circles), $(\gamma_{c}+\gamma_{X1})/2$ (purple left-triangle) in $\bold{a}$ and $\Omega_{2}$ (red square), $(\gamma_{c}+\gamma_{X2})/2$ (purple right-triangle) in $\bold{b}$. $\gamma_{c}=2.7 meV$ is constant with temperature. $\gamma_{X1}$ and $\gamma_{X2}$ are measured independently from bare hBL. $\Omega_{1}$ and $\Omega_{2}$ drop below the average linewidth at about 100K, showing the transition to the weak-coupling regime.}
    \label{fig:T-depen}
\end{figure*}
\begin{figure*}[t]
	\includegraphics[width=\linewidth]{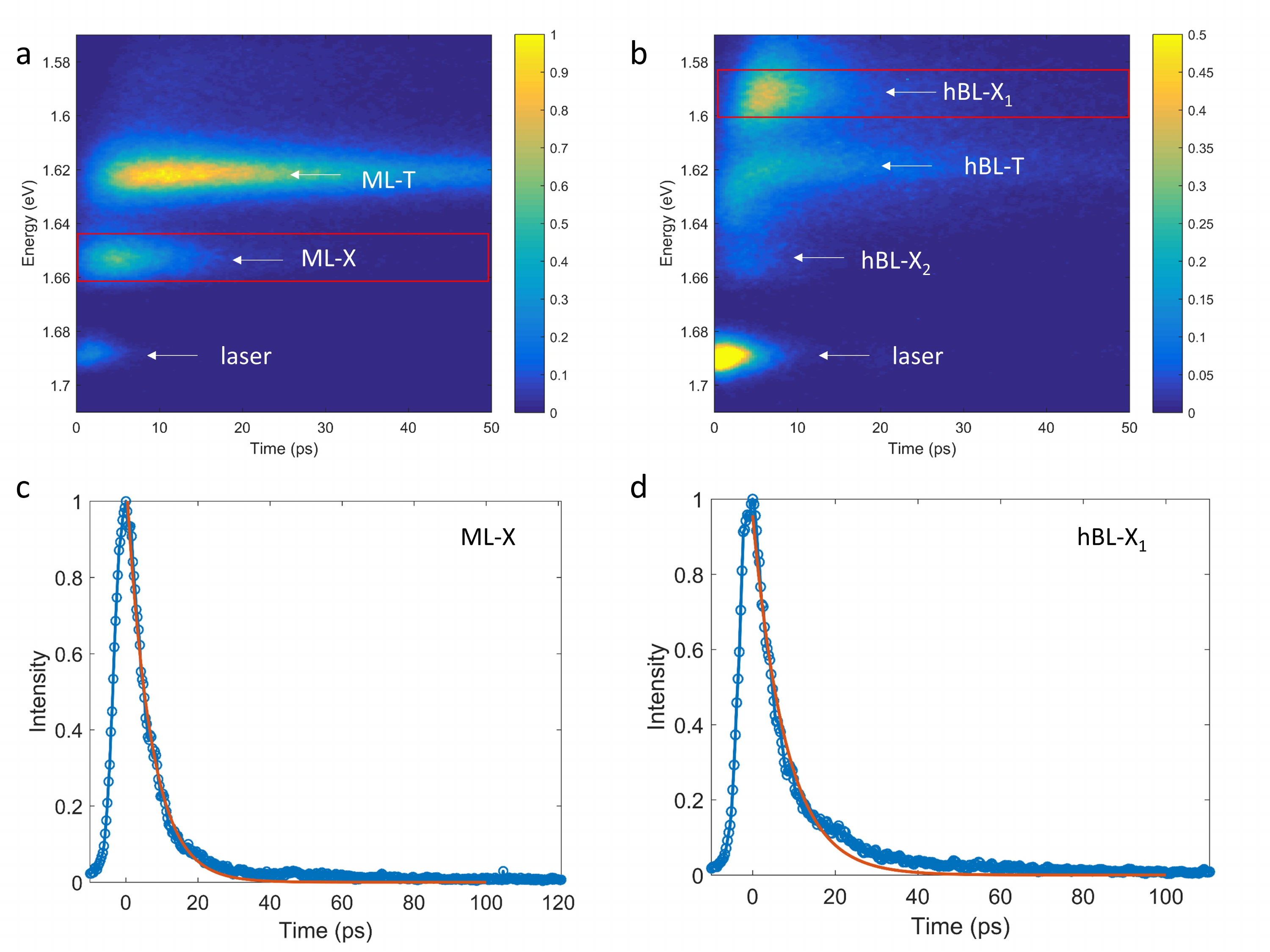}
	\caption{{\bfseries Time resolved photoluminescence (TRPL) of ML exciton, ML trion, hBL excitons, and hBL trion.} TRPL spectra measured by streak camera for ML \MoSe (ML) $\bold{a}$ and hBL \WS/\MoSe (hBL) $\bold{b}$. (The hBL data is collected from a different sample from the main text, which is not integrated with microcavity) The different resonances are labelled with white arrows including ML exciton (ML-X), ML trion (ML-T), \moire exciton at higher energy (hBL-X$_2$), \moire excion at lower energy (hBL-X$_1$), and \moire trion (hBL-T).   $\bold{c}$ and $\bold{d}$ shows the time resolved decay of ML-X and hBL-X$_1$ respectively by integrating the spectrum in the range labeled by the red rectangles in a and b.  The red solid lines in $\bold{c}$ and $\bold{d}$ are the fits with single exponential decay function. PL decay time for ML-X and hBL-X$_1$ are 6.7 ps and 8.0 ps respectively.}
    \label{fig:T-depen}
\end{figure*}
\begin{figure*}[t]
  \includegraphics[width=\linewidth]{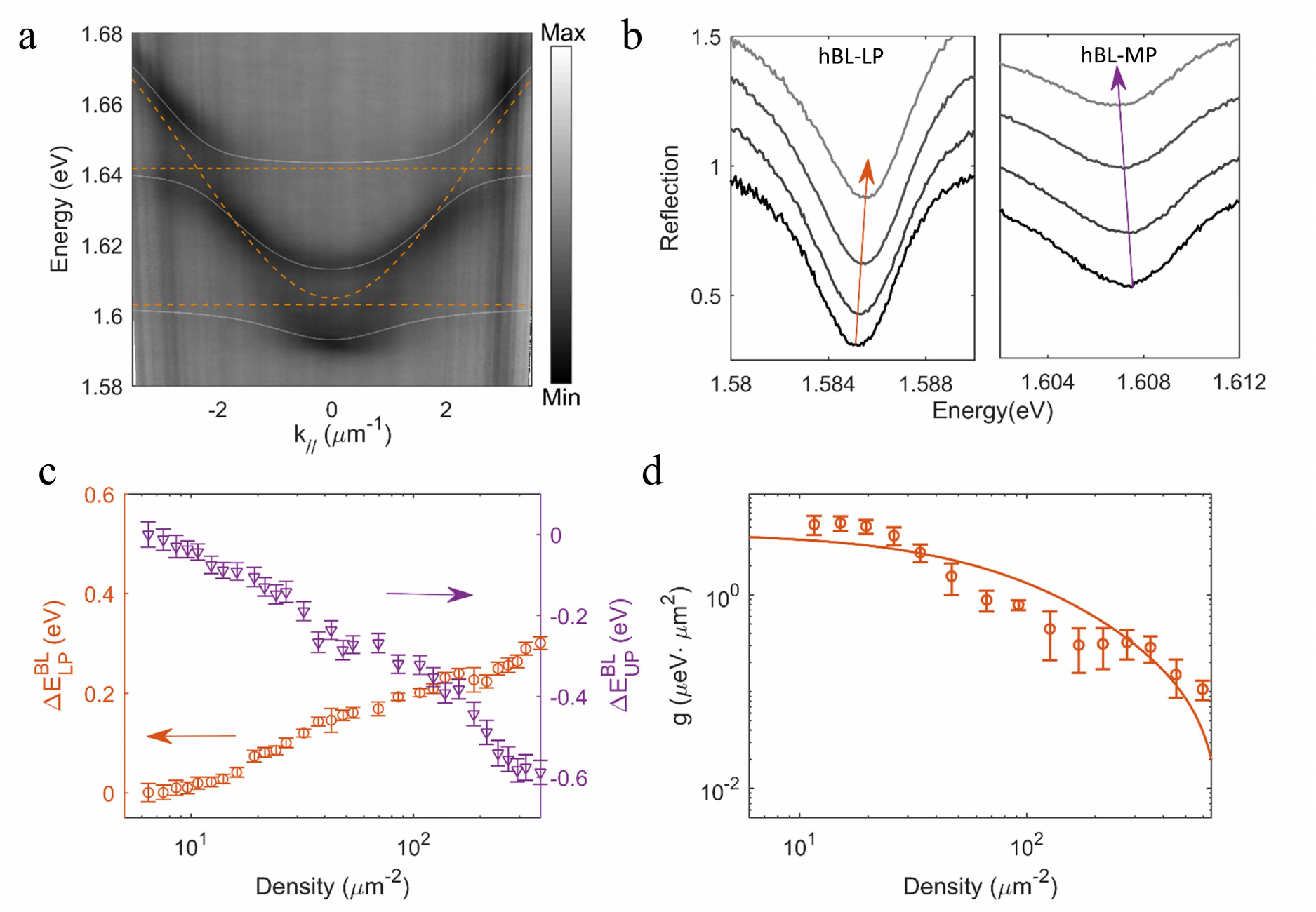}
	\caption{{\bfseries Strong nonlinearity measured in another device.} Measurement taken on a different sample shows reproducibility of the results. $\bold{a}$ Angle resolved white light reflection spectra taken at 5K on the second sample. White solid lines are the fits using coupled oscillator model. The dashed white lines are the fitted energies of the uncoupled cavity photon and excitons. $\bold{b}$ Power dependent reflection spectra for the lower polariton (left panel) and middle polariton (right panel).$\bold{c}$ Shift of polariton energies vs. the carrier density (log scale) obtained from $\bold{b}$. $\bold{d}$ Extracted nonlinear coefficients for lower polariton (red circles) and the calculations using fitted polariton energies (solid line).}
	\label{fig:sample2}
\end{figure*}
\begin{figure*}[t]
  \includegraphics[width=\linewidth]{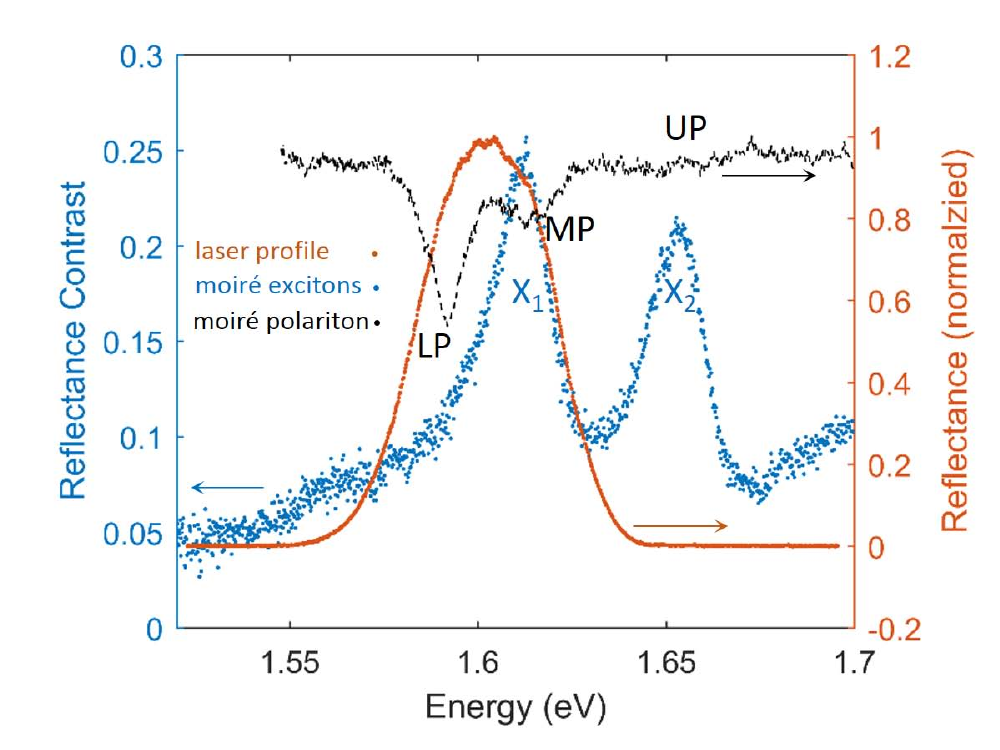}
	\caption{{\bfseries Profile of the laser used for nonlinearity characterization.} Profile of pulsed laser (red dot) used for the nonlinearity measurement of \moire polaritons, compared with the \moire exciton X$_1$ and X$_2$ (blue dots), and \moire polaritons (black dots). The laser has a negligibly small tail on X$_2$ and upper polariton.
}
	\label{fig:simulation}
\end{figure*}
\begin{figure*}[t]
	\includegraphics[width=0.6\linewidth]{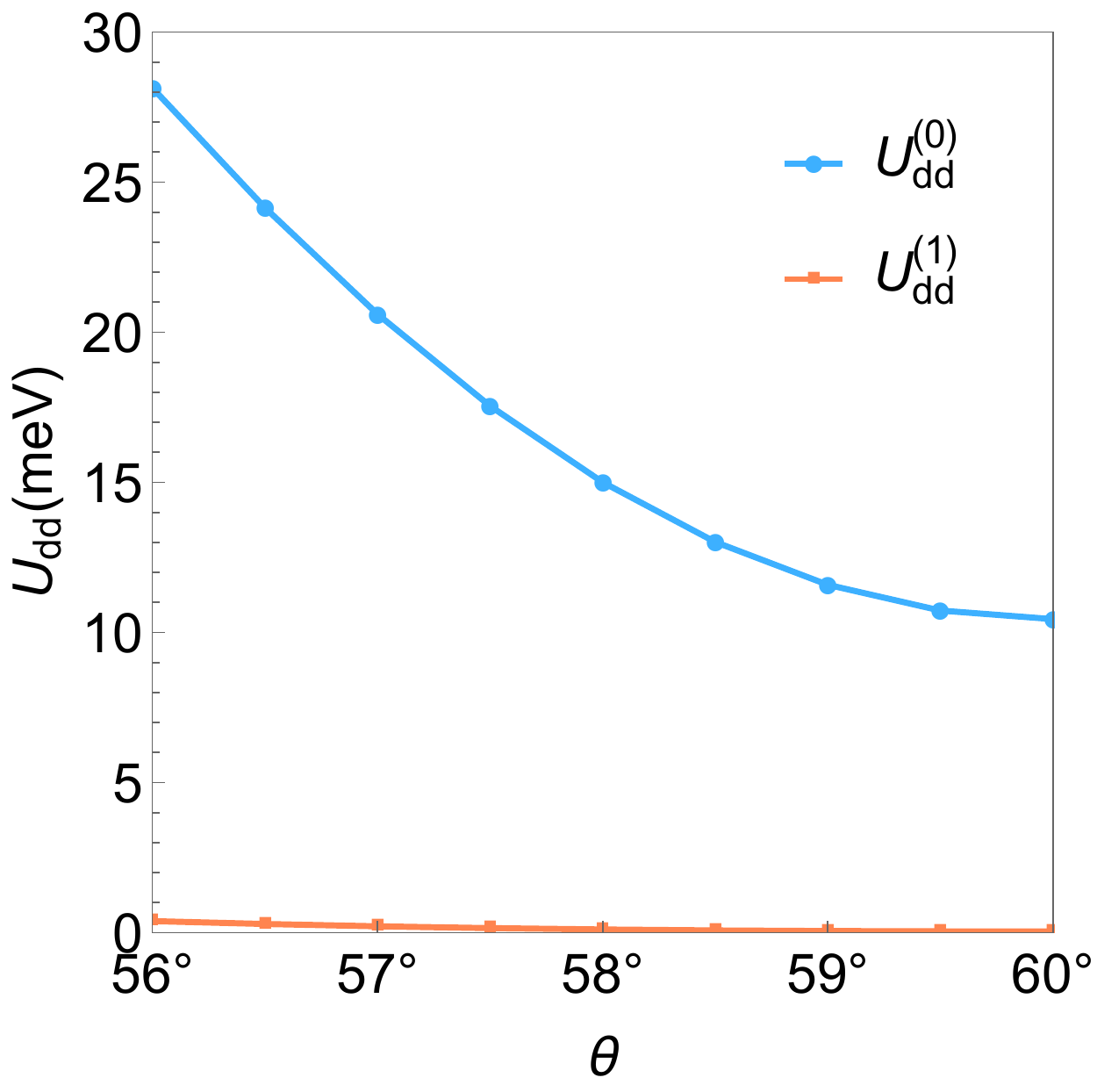}
	\caption{\bfseries Dipole-Dipole interaction strength as a function of twist angle $\theta$.} $U_{dd}^{(0)}$ and $U_{dd}^{(1)}$ are respectively onsite and nearest-neighbor interaction strength.
    \label{fig:Udd}
\end{figure*}
\begin{figure*}[t]
  \includegraphics[width=\linewidth]{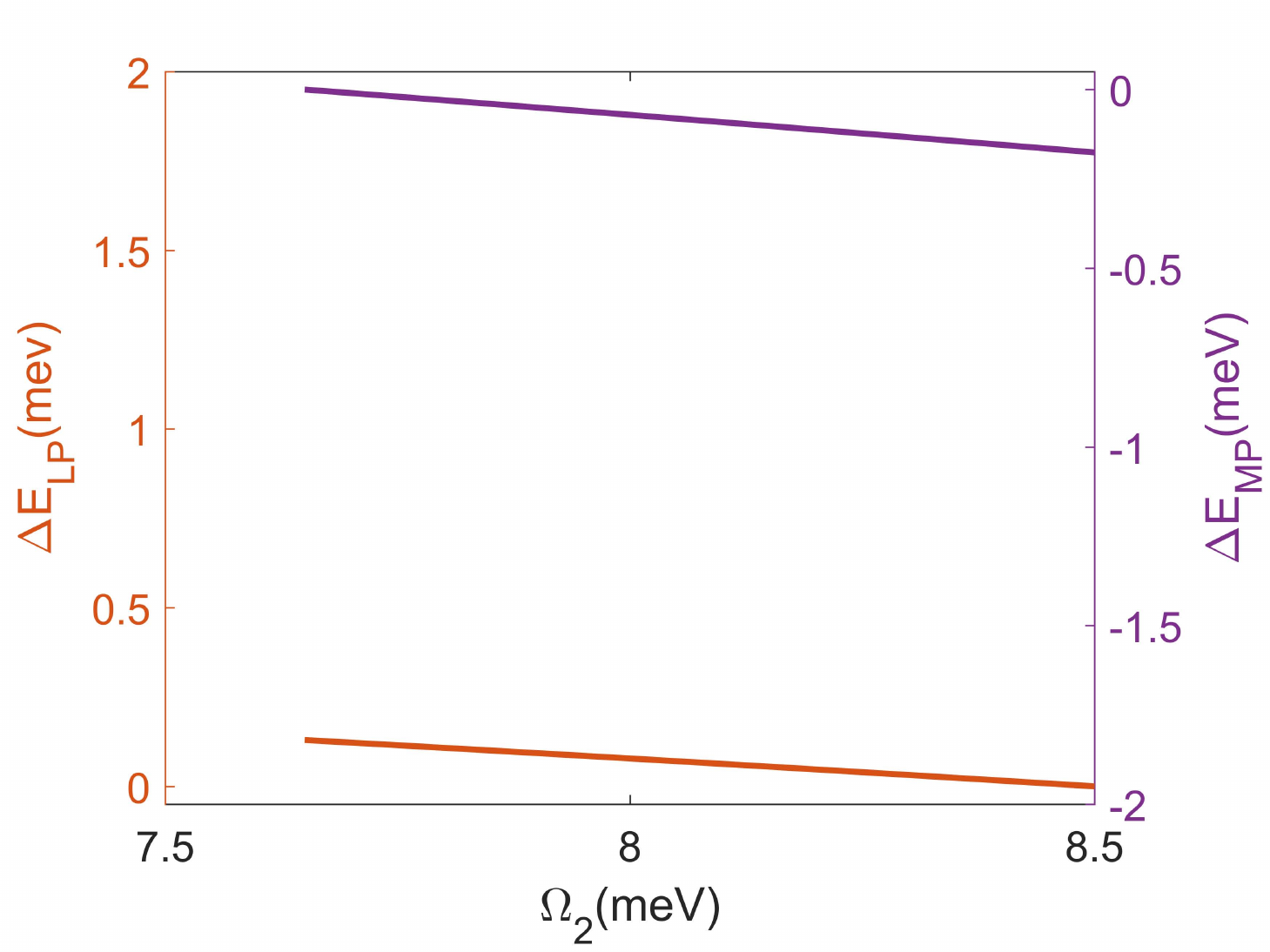}
	\caption{{\bfseries Effect on the nonlinearity from the \moire exciton X$_2$.} Energy shifts of $E^{BL}_{LP}$ and $E^{BL}_{MP}$ from the saturation of upper \moire exciton X$_2$. In the main test, we ignore the effects of \moire exciton X$_{2}$ on the nonlinearity of lower and middle polaritons. To quantitatively estimate the effects, we calculate the $\Delta E^{BL}_{LP}$ and $\Delta E^{BL}_{MP}$ as a function of coupling strength ($\Omega_2$) of X$_{2}$. When the $\Omega_2$ changes by $10\%$, the $E_{{LP}}$ and $E_{{MP}}$ will shift within 0.16 meV, which is less than $8\%$ of the shift observed from the experiments (2meV) (Fig.~\ref{fig:analysis}c in the main text). So the change of $E_{LP}$ and $E_{MP}$ induced by $\Omega_2$ can be safely ignored for simplicity.}
	\label{fig:simulation}
\end{figure*}

%\begin{figure*}[t]
%	\includegraphics[width=\linewidth]{SIFigure5.png}
%	\caption{Extracting the saturation density by fitting the density dependence of coupling strength of bilayer \moire $X_1$ at low and high density regions separately. The fitting function is $\Omega(n)={\Omega_{0}}/{\sqrt{1+\frac{n}{n_{s}}}}$. The yellow and red solid line are the fits at low and high density region respectively. And the corresponding fitted saturation densities are $n_{s}^{l}=(6.6\pm1.3)\times 10^2~\mu$m$^{-2}$ in the low density range of n$\leq$ 30 $\mu$m$^{-2}$ and $n_{s}^{h}=(6.2\pm0.3)\times 10^4~\mu$m$^{-2}$ in the high density range of n$\geq$ 500 $\mu$m$^{-2}$.}
%    \label{fig:saturation density}
%\end{figure*}
%

\end{document}